\documentclass[preprint,showpacs,preprintnumbers,amsmath,amssymb]{revtex4}
\usepackage{graphicx}% Include figure files
\usepackage{dcolumn}% Align table columns on decimal point
\usepackage{bm}% bold math
\newif\ifappendixon
\begin{document}

%\preprint{APS/123-QED}

\title{An exact solution of the slow-light problem}

\author{A.V. Rybin }
\affiliation{Department of Physics, University of Jyv\"askyl\"a PO
Box 35, FIN-40351
\\ Jyv{\"a}skyl{\"a}, Finland}
\email{andrei.rybin@phys.jyu.fi}

\author{ I.P. Vadeiko}
\affiliation{School of Physics and Astronomy, University of St
Andrews, North Haugh, St Andrews, KY16 9SS, Scotland}
\email{iv3@st-andrews.ac.uk}

\date{ October 7, 2004}

\author{ A. R. Bishop}
\affiliation{Theoretical Division and Center for Nonlinear
Studies, Los Alamos National Laboratory, Los Alamos, New Mexico
87545, USA} \email{arb@lanl.gov}

\begin{abstract}
We investigate  propagation of a slow-light soliton in atomic
vapors and Bose-Einstein condensates described by the nonlinear
$\Lambda$-model. We show that the group velocity of the soliton
monotonically decreases with the intensity of the controlling
laser field, which decays exponentially after the laser is
switched off. The shock wave of the vanishing controlling field
overtakes the slow soliton and stops it, while the optical
information is recorded in the medium in the form of spatially
localized polarization. We find an explicit exact solution
describing the whole process within the slowly varying amplitude
and phase approximation. Our results point to the possibility of
addressing spatially localized memory formations and moving these
memory bits along the medium in a controllable fashion.

\end{abstract}
%Use showkeys class option if keyword
\maketitle

\begin{figure}
\includegraphics[width=50mm]{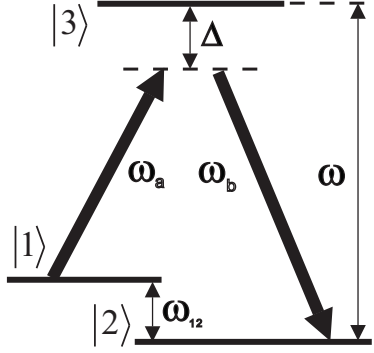}
\caption{\label{fig:spec1} The $\Lambda$-scheme for working energy
levels of alkali atoms.}
\end{figure}

In this paper we study the interaction of light with a gaseous
active medium whose working energy levels are well approximated by
the $\Lambda$-scheme. The model is a very close prototype for a gas
of alkali atoms, whose interaction with the light is approximated by
the structure of levels of the $\Lambda$-type given in
Fig.~\ref{fig:spec1}. We consider the case when the atoms are cooled
down to  microkelvin temperatures in order to suppress the Doppler
shift and increase the coherence life-time for the ground levels.
Typically, in experiments \cite{Hau:1999, Liu:2001, Phillips:2001,
Bajcsy:2003, Braje:2003, Mikhailov:2004} the pulses have the length
of microseconds, which is much shorter than the coherence life-time
and longer than the optical relaxation times. The gas cell is
illuminated by two circularly polarized optical beams co-propagating
in the $z$-direction. One $\sigma^-$-polarized field is denoted as
$a$, and the other $\sigma^+$-polarized field is denoted as $b$. We
study dynamics of the fields within the slowly varying amplitude and
phase approximation (SVEPA). The total superposition of the fields
is represented in the form
\begin{equation}\label{fields}
     \vec{E}=\vec{e}_a\, \mathcal{E}_a e^{i(k_a x-\omega_a t)} +
     \vec{e}_b\, \mathcal{E}_b e^{i(k_b x-\omega_b t)} +c.c.
\end{equation}
Here, $k_{a,b}$ are the wave numbers, while the vectors
$\vec{e}_a,\vec{e}_b$ describe polarizations of the fields with
the amplitudes $\mathcal{E}_{a,b}$. We introduce two corresponding
Rabi frequencies
\begin{equation}\label{Rabi_f}
    \Omega_a=\frac{2\mu_{a}\mathcal{E}_a}{\hbar},\;
    \Omega_b=\frac{2\mu_{b}\mathcal{E}_b}{\hbar},
\end{equation}
where $\mu_{a,b}$ are dipole moments of corresponding transitions
in the atom.

In the interaction picture and within the SVEPA  the Hamiltonian
describing the interaction of a three-level atom with the fields
assumes the form
\begin{eqnarray}
\label{H_lam} &H_\Lambda=H_0+H_I,\;H_0=-\frac\Delta2 D,\\
&H_I=-\frac12 \left({\Omega_a |3\rangle\langle1|
+\Omega_b|3\rangle\langle2|}\right) +h.c.,\nonumber
\end{eqnarray}
where
$$
D=\left(%
\begin{array}{ccc}
  1 & 0 & 0 \\
  0 & 1 & 0 \\
  0 & 0 & -1 \\
\end{array}%
\right).
$$
We set $\hbar=1$.

Dynamics of the fields is described by the Maxwell equations, which
within the SVEPA take the form
\begin{equation}\label{Maxwell_m}
    \partial_\zeta H_I=i\frac {\nu_0}4 \left[{D,\rho}\right].
\end{equation}
Here we have introduced new variables $\zeta=(z-z_0)/c$,
$\tau=t-(z-z_0)/c$, $\rho$ is the density matrix in the
interaction representation, $n_A$ is the density of atoms, and
$\epsilon_0$ is the vacuum susceptibility. For many experimental
situations it is typical that the coupling constants
$\nu_{a,b}=(n_A |\mu_{a,b}|^2 \omega_{a,b})/\epsilon_0$ are almost
the same. Therefore we assume that $\nu_a\approx\nu_b=\nu_0$.
Together with the Liouville equation,
\begin{equation}\label{Liouv}
    \partial_\tau \rho=-i\left[{H_\Lambda,\rho}\right],
\end{equation}
we obtain a system of equations
Eqs.(\ref{Maxwell_m}),(\ref{Liouv}), which is exactly solvable in
the framework of the Inverse Scattering (IS) method
\cite{fad,Park:1998,gab,Rybin:2004}.

At this point it is worth discussing the initial and boundary
conditions underlying the physical problem in question. We
consider a semi-infinite $\zeta\ge0$ active medium with a pulse of
light incident at the point $\zeta=0$ (initial condition). This
means that the evolution is considered with respect to the {\it
space} variable $\zeta$, while the boundary conditions should be
specified with respect to the  variable $\tau$. In this paper we
assume  the atom-field system to be prepared in an  initial state
corresponding to the typical experimental setup (see e.g.
\cite{Hau:1999,Liu:2001,Bajcsy:2003}):
\begin{equation}\label{init_fields0}
    \Omega^{(0)}_a=0,\; \Omega^{(0)}_b=\Omega(\tau),\;
    |\psi_{at}\rangle=e^{-i\frac\Delta2 \tau}|1\rangle.
\end{equation}
The state satisfies the Maxwell-Bloch system of equations
Eqs.(\ref{Maxwell_m}),(\ref{Liouv}). The field $\Omega(\tau)$
plays the role of the controlling background field generated by a
laser. In \cite{Rybin:2004} (see also~\cite{Park:1998}) for the
constant background field $\Omega(\tau)=\Omega_0=\Omega_0^\ast$ we
found the slow-light soliton
\begin{equation}\label{slow_const}\Omega_a=
\frac{-i\sqrt{2\varepsilon_0}\Omega_0}
{\sqrt{\varepsilon_0+\sqrt{\varepsilon_0^2-\Omega_0^2}}}\,
\mathrm{sech}\,\phi_s,\, \Omega_b=\Omega_0\tanh\phi_s,
\end{equation}
where $\phi_s=\frac{\nu_0\zeta}{2\varepsilon_0}
-\frac\tau2\left({\varepsilon_0-\sqrt{\varepsilon_0^2-\Omega_0^2}}\right)+\phi_0$.
In these expressions, for simplicity we set $\Delta=0$,
$\varepsilon_0>\Omega_0$. The meaning of the parameter
$\varepsilon_0$ is explained below. It can be readily seen that
the speed of the slow-light soliton sent into the system depends
on the intensity of the controlling background field. In the
simplifying approximation $\frac{
\Omega_0^2}{\varepsilon_0^2}\ll1$, the group velocity
reads~\cite{Rybin:2004}

\begin{equation}\label{speed_constant}
v_g\approx c\frac{\Omega_0^2}{2\nu_0}.
\end{equation}

This expression immediately  suggests  a  plausible conjecture that
when the controlling field is switched off the soliton stops
propagating while the information borne by the soliton remains in
the medium in the form of an imprinted polarization flip.  In what
follows we refer to this flip as   a memory cell or memory imprint.
For the time-dependent background field some approximate solutions
based on the methods of scaled time are studied in \cite{Grobe:1994,
Bajcsy:2003, Dey:2003, ulf:2004}.

In this paper we provide  strong evidence substantiating  the
dynamical mechanism formulated above as a hypothesis.   This
evidence is based on the exact solution described below. We
envisage the following dynamics scenario. We assume that the
slow-light soliton was created in the system before the moment of
time $t=0$ and is propagating on the background of the constant
controlling field $\Omega_0$. This is the property of the soliton,
supported by the exact solution~\cite{Rybin:2004} that, after the
soliton has passed, the medium returns to the initial quantum
state Eq.(\ref{init_fields0}). Suppose that at the moment in time
$t=0$ the laser source of the controlling field is switched off.
We assume that after this moment the background field will
exponentially decay with some characteristic rate $\alpha$. The
exponential front of the vanishing controlling field will then
propagate into the medium, starting from the point $z=z_0$, where
the laser is placed. The state of the quantum system
Eq.(\ref{init_fields0}) is dark for the controlling field.
Therefore the medium is transparent for the spreading front of the
vanishing field, which then propagates  with the speed of light,
eventually overtaking the slow-light soliton and stopping it.

To realize the mechanism described above we define the time
dependence of $\Omega(\tau)$ as follows:
\begin{equation}\label{bg_field}
    \Omega(\tau)=\left\{ {\begin{array}{ll}
                            \Omega_0, &  \tau<0\\
                            \Omega_0 \exp(-\alpha\, \tau), & \tau\geq 0 \\
                          \end{array}}\right. .
\end{equation}
This regime of switching the field off is quite realistic. An
experimental setup, where the parameter $\alpha$ becomes
experimentally adjustable can be easily envisaged. Greater values
of $\alpha$ correspond to steeper fronts of the incoming
background wave. In what follows we also discuss this limit.

Using the methods of our previous work~\cite{Rybin:2004}, we
construct the exact analytical single-soliton solution of the
problem. For the fields the solutions are
\begin{eqnarray}\label{fields1}
\tilde\Omega_a&=&\frac{(\lambda^*-\lambda)w(\tau,\lambda)
} {\sqrt{1+|w(\tau,\lambda)|^2}}\; e^{i\tilde\theta_s}\, \mathrm{sech}\tilde\phi_s,\\
\tilde\Omega_b&=&\frac{(\lambda-\lambda^*)w(\tau,\lambda) }
{1+|w(\tau,\lambda)|^2}\,e^{\tilde\phi_s}\,
\mathrm{sech}\tilde\phi_s-\Omega(\tau),\nonumber
\end{eqnarray}
and for the atomic medium
\begin{eqnarray}\label{atoms1}
|\tilde\psi_{at}\rangle=e^{-i\frac\Delta2
\tau}\left[{\left({\frac{\lambda^*-\Delta}{|\lambda-\Delta|}-
\frac{\tilde\Omega_a
e^{i\left({\frac{\nu_0\zeta}{2(\lambda-\Delta)}-\tilde\theta_0}\right)-z(\tau,\lambda)}}
{2|\lambda-\Delta|w(\tau,\lambda)} }\right)|1\rangle+
\frac{\tilde\Omega_a} {2|\lambda-\Delta|w(\tau,\lambda)}
|2\rangle-\frac{\tilde\Omega_a} {2|\lambda-\Delta|}
|3\rangle}\right].\quad
\end{eqnarray}
 The  parameters of
the slow-light soliton are
\begin{eqnarray}\label{sigma_params}
\tilde\phi_s&=&\tilde\phi_0-
\frac{\nu_0\mathrm{Im}(\lambda)\zeta}{2|\Delta-\lambda|^2}+
\mathrm{Re}(z(\tau,\lambda))+\frac12\ln({1+|w(\tau,\lambda)|^2}),\nonumber\\
\tilde\theta_s&=&\tilde\theta_0-\frac{\nu_0\zeta}2
\mathrm{Re}\left({
\frac1{\lambda-\Delta}}\right)+\mathrm{Im}(z(\tau,\lambda)),\,\,
\lambda\in {\mathbb C}.
\end{eqnarray}
We find the functions $w,z$. For $\tau<0$ these functions are:
\begin{eqnarray}\label{w0_z0}
   w\equiv w_0=\frac{\Omega_0}{\lambda+\sqrt{\lambda^2+{\Omega_0}\!^2}},\;
z(\tau,\lambda)=\frac i2 \Omega_0 w_0 \tau,
\end{eqnarray}
while for $\tau\geq 0$ the functions $w,z$ take the form:
\begin{eqnarray}
\label{z} &z(\tau,\lambda)&=-\alpha\gamma\tau+\ln \frac{{{\cal C}}
J_{-\gamma} \left(-\frac{\Omega(\tau)}{2\alpha}\right)+ J_{\gamma}
\left(-\frac{\Omega(\tau)}{2\alpha}\right)}{{\cal C} J_{-\gamma}
(-\frac{\Omega_0}{2\alpha})+
J_{\gamma} \left(-\frac{\Omega_0}{2\alpha}\right)},\quad\quad\\
\label{w}  &w(\tau,\lambda)&=i\frac{{{\cal C}} J_{1-\gamma}
\left(-\frac{\Omega(\tau)}{2\alpha}\right)- J_{\gamma-1}
\left(-\frac{\Omega(\tau)}{2\alpha}\right)}{{\cal C} J_{-\gamma}
\left(-\frac{\Omega(\tau)}{2\alpha}\right)+ J_{\gamma}
\left(-\frac{\Omega(\tau)}{2\alpha}\right)},
\end{eqnarray}
where $\gamma=\frac{\alpha+i\lambda}{2\alpha}$ and $J_\nu$ are
Bessel functions. The constant ${\cal C}$ is uniquely defined by
the condition $w(0,\lambda)=w_0$
\begin{equation}\label{const_C}
    {\cal C}=\frac{-i w_0
J_{\gamma} (-\frac{\Omega_0}{2\alpha})+ J_{\gamma-1}
(-\frac{\Omega_0}{2\alpha})}{ J_{1-\gamma}
(-\frac{\Omega_0}{2\alpha})+ i w_0 J_{-\gamma}
(-\frac{\Omega_0}{2\alpha})}.
\end{equation}
To identify a physically relevant solution   we require that
$w(\infty,\lambda)=0$. This requirement places a restriction on
the parameter $\lambda$, such that $\mathrm{Im}(\lambda)<0$ and
hence $\mathrm{Re}(\gamma)>1/2$.

We now  discuss the physical properties of the exact solution
given above.   The group velocity of the slow-light soliton reads
\begin{equation}\label{speed1}
    \frac{v_g}c=\frac{|w(\tau,\lambda)|^2}
    {\frac{\nu_0(1+|w(\tau,\lambda)|^2)}{2|\Delta-\lambda|^2}+|w(\tau,\lambda)|^2}.
\end{equation}
Notice that in the case of the constant background field, i.e. in
the case $\alpha=0$, the conventional expressions for the
slow-light soliton Eq.(\ref{slow_const}) along with the expression
for the group velocity Eq.(\ref{speed_constant})  - the main
motivational quantity for this paper -  can be readily recovered
from Eqs.(\ref{fields1}),(\ref{speed1}).

We calculate the distance ${\cal L}_s(\alpha)$ that the slow-light
soliton will propagate from the moment $t=0$, when the laser is
switched off, until the full stopping of the signal. This distance
is
\begin{eqnarray}\label{delta_L}
  {\cal L}_s(\alpha)=\frac{2c|\Delta-\lambda|^2}{\nu_0\mathrm{Im}(\lambda)}
  \,\tilde\phi_s|_{\tau=0}^{\tau=\infty}= \frac{2c|\Delta-\lambda|^2}{\nu_0|\mathrm{Im}(\lambda)|}
\left[\frac12\ln\left(1+|w_0|^2\right)-\mathrm{Re}(z(\infty,\lambda))\right].
\end{eqnarray}
It is clear that
$$
z(\infty,\lambda)=\ln \frac{{\cal C}
\left(-\frac{\Omega_0}{4\alpha}\right)^{-\gamma}/\mathrm{\Gamma}(1-\gamma)}{{\cal
C} J_{-\gamma} (-\frac{\Omega_0}{2\alpha})+ J_{\gamma}
\left(-\frac{\Omega_0}{2\alpha}\right)}.$$  Notice that
$\mathrm{Re}\,z(\infty,\lambda)\leq0$ and hence  ${\cal
L}_s(\alpha)>0$. The case when the controlling field is instantly
switched off corresponds to the limit $\alpha\rightarrow\infty$.
In this limit the profile of the background field approaches the
Heaviside step-function, and we find that
$\lim\limits_{\alpha\to\infty}\mathrm{Re}\,z(\infty,\lambda)=0$.
In this case, as  is intuitively evident, the soliton will still
propagate over some finite distance. Notice  that the limits
$\tau\rightarrow\infty$ and $\alpha\rightarrow\infty$ do not
commute and the latter should be taken after the former.

The half-width of the polarization flip  written into the medium
after the soliton is completely stopped is
\begin{equation}\label{width_P}
{\cal W}_s= 4c\ln(2+\sqrt3)
    \frac{|\Delta-\lambda|^2}{\nu_0\left|\mathrm{Im}(\lambda)\right|}.
\end{equation}
It is important to notice that the width  Eq.(\ref{width_P}) of the
imprinted memory cells do not depend on the  rate $\alpha$. In other
words, the width of the memory cell is not sensitive to how rapidly
the controlling field is switched off. This leads  to an important
conclusion. Indeed, through the variation of the experimentally
adjustable parameter $\alpha$  it is possible to control the
location of the memory cell, while the characteristic size of the
imprinted memory cell remains intact. This result is strongly
supported by recent experiments \cite{Dutton:2004}.

\begin{figure}
\includegraphics[width=60mm]{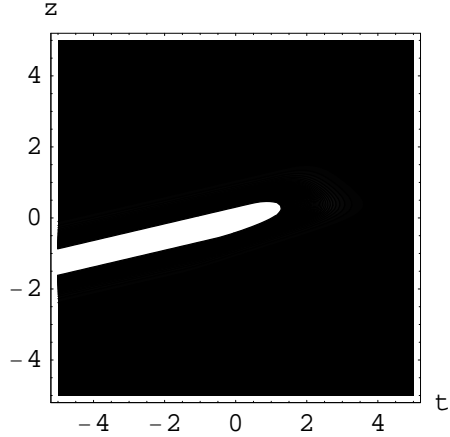}
\caption{\label{fig:spec2} The slow-light soliton is stopped
\cite{Bajcsy:2003}. Contour plot of the intensity $I_a$ of the field
$\tilde \Omega_a$. The parameters used for all the plots in this
report are the same: $\Delta=0$, $\Omega_0=2$, $\varepsilon_0=2.1$,
$\nu_0=10$, $\alpha=1$.}
\end{figure}

\begin{figure}
\includegraphics[width=60mm]{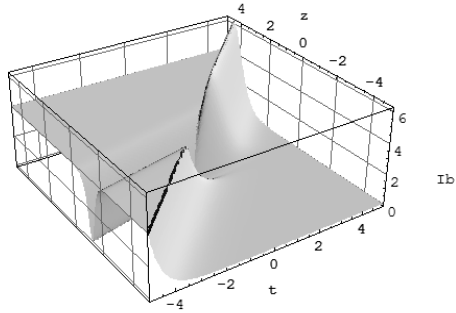}
\caption{\label{fig:spec3} The intensity $I_b$ of the field
$\tilde \Omega_b$ as a function of time $t$ and space $z$. The
parameters are defined in Fig.~\ref{fig:spec2}. The groove
corresponds to the slow-light  soliton.  The soliton collides with
the shock wave of the vanishing control field, whose  front has an
exponential profile. }
\end{figure}

\begin{figure}
\includegraphics[width=80mm]{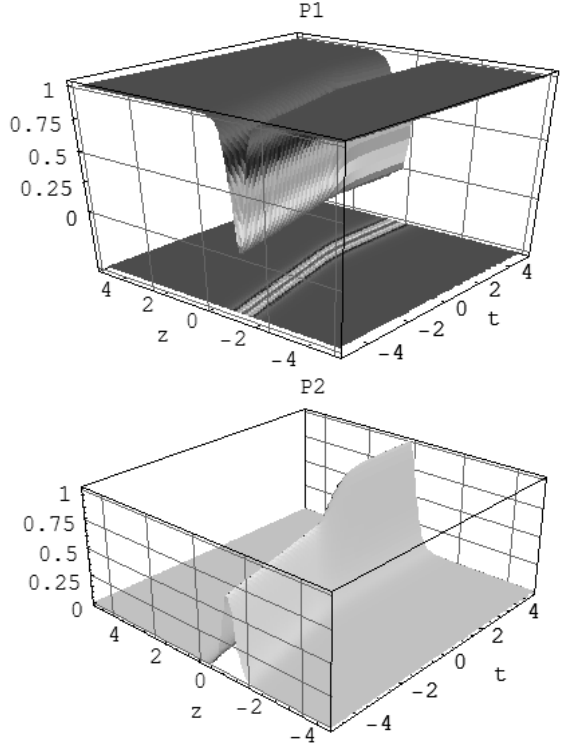}
\caption{\label{fig:spec4} Imprinting the information. We plot the
dynamics of the population $P_1$ of the level $|1\rangle$ and
$P_2$ of the level $|2\rangle$ as functions of time $t$ and space
$z$. It is clear that after the collision the populations freeze.}
\end{figure}

In what follows we assume the parameter $\lambda$ to be imaginary,
i.e. $\lambda=-i \varepsilon_0$, with $\varepsilon_0>\Omega_0$. We
demonstrate the slow-light dynamics in  Fig.~\ref{fig:spec2},
where we show how the slow-light soliton stops and disappears.
In  Fig.~\ref{fig:spec3} we demonstrate the behavior of the
intensity in the channel $b$. The groove in the constant
background field corresponds to the slow-light soliton. This
groove is complementary to the peak in the channel $a$. The shock
wave, whose front has an exponential profile, propagates  with the
speed of light, reaches the slow-light soliton and stops it.
Notice that after the collision with the slow-light soliton the
front of the shock wave shows some short peak at a level higher
than the background intensity. This effect reflects an essentially
nonlinear nature of interactions inherent to the considered
dynamical system.
In  Fig.~\ref{fig:spec4} we plot the dynamics of the polarization
flip in the atomic medium, which bears the information and stores
it in the medium. Before the shock wave has approached, the
spatially localized population of the level $|2\rangle$ moves
together with the slow-light soliton as a composite whole. This is
a nonlinear analog of the dark-state polariton
\cite{Fleischhauer:2000}. After the shock wave has arrived the
population is frozen at the position where it was hit by the shock
wave.

It is important to notice that, before the slow-light soliton is
stopped, the population on the level $|2\rangle$ does not reach
unity, because there is a remnant population in the upper state
$|3\rangle$. However, after the polariton is completely stopped, the
level $|3\rangle$ depopulates, while the population of the level
$|2\rangle$ reaches unity at the maximum of the signal. Before and
after the collision, the population of the level $|1\rangle$ at the
minimum of the groove always vanishes. The result is valid for zero
detuning. From our solution Eq.(\ref{atoms1}) it is not difficult to
estimate the maximum population of the level $|2\rangle$ for finite
detuning. After the soliton is completely stopped the population at
the maximum is $ {\varepsilon_0^2}/{(\varepsilon_0^2+\Delta^2)}$.

Any destructive influence of the relaxation processes  on the
overall picture of dynamics described above is negligible. Indeed,
as we show in \cite{Rybin:2004}, the population of the upper level
$|3\rangle$ is proportional to the intensity of the background
field $\Omega_0$, which is required to be small, to ensure a small
velocity of the signal (cf. Eq.(\ref{speed_constant})).

{\em Discussion}. In this report we have  investigated the dynamics
of a slow-light soliton whose group velocity explicitly depends on
the background (controlling) field. Taking advantage of an explicit
exactly solvable example, we demonstrate that the soliton can indeed
be stopped, provided that the background field vanishes. After the
signal is stopped, the information borne by the soliton is imprinted
into the medium in the form of a spatially localized polarization
flip.  The width of the imprinted memory, in our example, does not
depend on how rapidly the field vanishes. However, the position of
the spatially localized optical memory imprint is controlled by the
experimentally adjustable parameter $\alpha$. Therefore, we have
shown  that our approach allows addressing of optical memory
recorded into an atomic medium at an exact location without changing
the characteristic size of the spatial domain occupied by the
memory. The imprinted memory can be subsequently read. Moreover, our
results point to the possibility of switching the controlling field
on and off repeatedly, and by virtue of this mechanism {\em move}
the imprinted memory through the medium in a strictly controlled
fashion. The localized polarization will follow the sequence of
step-like laser pulses and therefore hop over finite distances in a
prescribed manner.

Another experimentally interesting application of our results  is to
inject multi-soliton packets into the medium, creating solitonic
'trains', whose 'carriages' are subsequently stopped  by the shock
wave of the vanishing background. In this way a multi-bit optical
memory can be realized.
%These and other interesting ramifications
%\cite{Bullough:2003} of the results reported in this paper will be
%discussed elsewhere .

IV acknowledges the support of the Engineering and Physical Sciences
Research Council, United Kingdom. %AR acknowledges useful discusions
%with R.K. Bullough on quantum memory problem.
Work at Los Alamos National Laboratory is supported by the USDoE.
\bibliography{paper}

\begin{thebibliography}{15}
\expandafter\ifx\csname natexlab\endcsname\relax\def\natexlab#1{#1}\fi
\expandafter\ifx\csname bibnamefont\endcsname\relax
  \def\bibnamefont#1{#1}\fi
\expandafter\ifx\csname bibfnamefont\endcsname\relax
  \def\bibfnamefont#1{#1}\fi
\expandafter\ifx\csname citenamefont\endcsname\relax
  \def\citenamefont#1{#1}\fi
\expandafter\ifx\csname url\endcsname\relax
  \def\url#1{\texttt{#1}}\fi
\expandafter\ifx\csname urlprefix\endcsname\relax\def\urlprefix{URL }\fi
\providecommand{\bibinfo}[2]{#2}
\providecommand{\eprint}[2][]{\url{#2}}

\bibitem[{\citenamefont{Hau et~al.}(1999)\citenamefont{Hau, Harris, Dutton, and
  Behroozi}}]{Hau:1999}
\bibinfo{author}{\bibfnamefont{L.~N.} \bibnamefont{Hau}},
  \bibinfo{author}{\bibfnamefont{S.~E.} \bibnamefont{Harris}},
  \bibinfo{author}{\bibfnamefont{Z.}~\bibnamefont{Dutton}}, \bibnamefont{and}
  \bibinfo{author}{\bibfnamefont{C.~H.} \bibnamefont{Behroozi}},
  \bibinfo{journal}{Lett.\ to Nature} \textbf{\bibinfo{volume}{397}},
  \bibinfo{pages}{594} (\bibinfo{year}{1999}).

\bibitem[{\citenamefont{Liu et~al.}(2001)\citenamefont{Liu, Dutton, Behroozi,
  and Hau}}]{Liu:2001}
\bibinfo{author}{\bibfnamefont{C.}~\bibnamefont{Liu}},
  \bibinfo{author}{\bibfnamefont{Z.}~\bibnamefont{Dutton}},
  \bibinfo{author}{\bibfnamefont{C.~H.} \bibnamefont{Behroozi}},
  \bibnamefont{and} \bibinfo{author}{\bibfnamefont{L.~V.} \bibnamefont{Hau}},
  \bibinfo{journal}{Lett.\ to Nature} \textbf{\bibinfo{volume}{409}},
  \bibinfo{pages}{490} (\bibinfo{year}{2001}).

\bibitem[{\citenamefont{Phillips et~al.}(2001)\citenamefont{Phillips,
  Fleischhauer, Mair, Walsworth, and Lukin}}]{Phillips:2001}
\bibinfo{author}{\bibfnamefont{D.~F.} \bibnamefont{Phillips}},
  \bibinfo{author}{\bibfnamefont{A.}~\bibnamefont{Fleischhauer}},
  \bibinfo{author}{\bibfnamefont{A.}~\bibnamefont{Mair}},
  \bibinfo{author}{\bibfnamefont{R.~L.} \bibnamefont{Walsworth}},
  \bibnamefont{and} \bibinfo{author}{\bibfnamefont{M.~D.} \bibnamefont{Lukin}},
  \bibinfo{journal}{Phys. Rev. Lett.} \textbf{\bibinfo{volume}{86}},
  \bibinfo{pages}{783} (\bibinfo{year}{2001}).

\bibitem[{\citenamefont{Bajcsy et~al.}(2003)\citenamefont{Bajcsy, Zibrov, and
  Lukin}}]{Bajcsy:2003}
\bibinfo{author}{\bibfnamefont{M.}~\bibnamefont{Bajcsy}},
  \bibinfo{author}{\bibfnamefont{A.~S.} \bibnamefont{Zibrov}},
  \bibnamefont{and} \bibinfo{author}{\bibfnamefont{M.~D.} \bibnamefont{Lukin}},
  \bibinfo{journal}{Lett.\ to Nature} \textbf{\bibinfo{volume}{426}},
  \bibinfo{pages}{638} (\bibinfo{year}{2003}).

\bibitem[{\citenamefont{Braje et~al.}(2003)\citenamefont{Braje, Balic, Yin, and
  Harris}}]{Braje:2003}
\bibinfo{author}{\bibfnamefont{D.~A.} \bibnamefont{Braje}},
  \bibinfo{author}{\bibfnamefont{V.}~\bibnamefont{Balic}},
  \bibinfo{author}{\bibfnamefont{G.~Y.} \bibnamefont{Yin}}, \bibnamefont{and}
  \bibinfo{author}{\bibfnamefont{S.~E.} \bibnamefont{Harris}},
  \bibinfo{journal}{Phys. Rev. A} \textbf{\bibinfo{volume}{68}},
  \bibinfo{pages}{041801(R)} (\bibinfo{year}{2003}).

\bibitem[{\citenamefont{Mikhailov et~al.}(2004)\citenamefont{Mikhailov,
  Sautenkov, Rostovtsev, and Welch}}]{Mikhailov:2004}
\bibinfo{author}{\bibfnamefont{E.~E.} \bibnamefont{Mikhailov}},
  \bibinfo{author}{\bibfnamefont{V.~A.} \bibnamefont{Sautenkov}},
  \bibinfo{author}{\bibfnamefont{Y.~V.} \bibnamefont{Rostovtsev}},
  \bibnamefont{and} \bibinfo{author}{\bibfnamefont{G.~R.} \bibnamefont{Welch}},
  \bibinfo{journal}{J. Opt. Soc. Am. B} \textbf{\bibinfo{volume}{21}},
  \bibinfo{pages}{425} (\bibinfo{year}{2004}).

\bibitem[{\citenamefont{Faddeev and Takhtadjan}(1987)}]{fad}
\bibinfo{author}{\bibfnamefont{L.~D.} \bibnamefont{Faddeev}} \bibnamefont{and}
  \bibinfo{author}{\bibfnamefont{L.~A.} \bibnamefont{Takhtadjan}},
  \emph{\bibinfo{title}{Hamiltonian Methods in the Theory of Solitons}}
  (\bibinfo{publisher}{Springer, Berlin}, \bibinfo{year}{1987}).

\bibitem[{\citenamefont{Park and Shin}(1998)}]{Park:1998}
\bibinfo{author}{\bibfnamefont{Q.-H.} \bibnamefont{Park}} \bibnamefont{and}
  \bibinfo{author}{\bibfnamefont{H.~J.} \bibnamefont{Shin}},
  \bibinfo{journal}{Phys.\ Rev.\ A} \textbf{\bibinfo{volume}{57}},
  \bibinfo{pages}{4643} (\bibinfo{year}{1998}).

\bibitem[{\citenamefont{Byrne et~al.}(2003)\citenamefont{Byrne, Gabitov, and
  Kova\v{c}i\v{c}}}]{gab}
\bibinfo{author}{\bibfnamefont{J.~A.} \bibnamefont{Byrne}},
  \bibinfo{author}{\bibfnamefont{I.~R.} \bibnamefont{Gabitov}},
  \bibnamefont{and}
  \bibinfo{author}{\bibfnamefont{G.}~\bibnamefont{Kova\v{c}i\v{c}}},
  \bibinfo{journal}{Physica D} \textbf{\bibinfo{volume}{186}},
  \bibinfo{pages}{69} (\bibinfo{year}{2003}).

\bibitem[{\citenamefont{Rybin and Vadeiko}(2004)}]{Rybin:2004}
\bibinfo{author}{\bibfnamefont{A.~V.} \bibnamefont{Rybin}} \bibnamefont{and}
  \bibinfo{author}{\bibfnamefont{I.~P.} \bibnamefont{Vadeiko}},
  \bibinfo{journal}{Journal of Optics B: Quantum and Semiclassical Optics}
  \textbf{\bibinfo{volume}{6}}, \bibinfo{pages}{416} (\bibinfo{year}{2004}).

\bibitem[{\citenamefont{Grobe et~al.}(1994)\citenamefont{Grobe, Hioe, and
  Eberly}}]{Grobe:1994}
\bibinfo{author}{\bibfnamefont{R.}~\bibnamefont{Grobe}},
  \bibinfo{author}{\bibfnamefont{F.~T.} \bibnamefont{Hioe}}, \bibnamefont{and}
  \bibinfo{author}{\bibfnamefont{J.~H.} \bibnamefont{Eberly}},
  \bibinfo{journal}{Phys.\ Rev.\ Lett.} \textbf{\bibinfo{volume}{73}},
  \bibinfo{pages}{3183} (\bibinfo{year}{1994}).

\bibitem[{\citenamefont{Dey and Agarwal}(2003)}]{Dey:2003}
\bibinfo{author}{\bibfnamefont{T.~N.} \bibnamefont{Dey}} \bibnamefont{and}
  \bibinfo{author}{\bibfnamefont{G.~S.} \bibnamefont{Agarwal}},
  \bibinfo{journal}{Phys. Rev. A} \textbf{\bibinfo{volume}{67}},
  \bibinfo{pages}{033813} (\bibinfo{year}{2003}).

\bibitem[{\citenamefont{Leonhardt}(2004)}]{ulf:2004}
\bibinfo{author}{\bibfnamefont{U.}~\bibnamefont{Leonhardt}},
  \bibinfo{journal}{Preprint ArXiv: quant-ph/0408046}  (\bibinfo{year}{2004}).

\bibitem[{\citenamefont{Dutton and Hau}(2004)}]{Dutton:2004}
\bibinfo{author}{\bibfnamefont{Z.}~\bibnamefont{Dutton}} \bibnamefont{and}
  \bibinfo{author}{\bibfnamefont{L.~V.} \bibnamefont{Hau}},
  \bibinfo{journal}{Phys. Rev. A} \textbf{\bibinfo{volume}{70}},
  \bibinfo{pages}{053831} (\bibinfo{year}{2004}).

\bibitem[{\citenamefont{Fleischhauer and Lukin}(2000)}]{Fleischhauer:2000}
\bibinfo{author}{\bibfnamefont{M.}~\bibnamefont{Fleischhauer}}
  \bibnamefont{and} \bibinfo{author}{\bibfnamefont{M.~D.} \bibnamefont{Lukin}},
  \bibinfo{journal}{Phys.\ Rev.\ Lett.} \textbf{\bibinfo{volume}{84}},
  \bibinfo{pages}{5094} (\bibinfo{year}{2000}).

\end{thebibliography}
\end{document}